\documentclass[a4paper,fleqn,usenatbib]{mnras}

\usepackage{amssymb,amsmath} 
\usepackage{txfonts}         
\usepackage[T1]{fontenc}
\usepackage{ae,aecompl}

\usepackage{graphicx}	
\usepackage{natbib}
\usepackage{multirow}

\title[Magnetic hot supergiants]{Discovery of magnetic A supergiants: the
descendants of magnetic main sequence B stars\thanks{Based
on observations obtained at the Canada-France-Hawaii Telescope (CFHT) operated
by the National Research Council of Canada, the Institut National des Sciences
de l'Univers of the CNRS of France, and the University of Hawaii, and at the
European Southern Observatory (ESO), Chile (program ID 094.D-0274A, 094.D-0274B,
and 095.D-0155A).}}

\author[C. Neiner et al.]{
Coralie Neiner$^{1}$\thanks{E-mail: coralie.neiner@obspm.fr}, 
Mary E. Oksala$^{2,1}$, 
Cyril Georgy$^{3}$, 
Norbert Przybilla$^{4}$, 
\newauthor
St\'ephane Mathis$^{5,1}$,
Gregg Wade$^{6}$,
Matthias Kondrak$^{4}$, 
Luca Fossati$^{7}$,
\newauthor
Aurore Blaz\`ere$^{8,1}$, 
Bram Buysschaert$^{1,9}$, 
and Jason Grunhut$^{10}$,
\\
$^{1}$LESIA, Observatoire de Paris, PSL Research University, CNRS, Sorbonne Universit\'es, UPMC Univ. Paris 06, Univ. Paris\\
Diderot, Sorbonne Paris Cit\'e, 5 place Jules Janssen, 92195 Meudon, France\\
$^{2}$Department of Physics, California Lutheran University, 60 West Olsen Road \#3700, Thousand Oaks, CA 91360, USA\\
$^{3}$Geneva Observatory, University of Geneva, chemin des Maillettes 51, 1290 Sauverny, Switzerland\\
$^{4}$Institut f\"ur Astro- und Teilchenphysik, Universit\"at Innsbruck, Technikerstr. 25/8, 6020, Innsbruck, Austria\\
$^{5}$Laboratoire AIM Paris-Saclay, CEA/DRF - CNRS - Universit\'e Paris Diderot, IRFU/DAp Centre de Saclay, 91191 Gif-sur-Yvette,\\
France\\
$^{6}$Department of Physics, Royal Military College of Canada, PO Box 17000 Kingston, ON K7K 7B4, Canada\\
$^{7}$Space Research Institute, Austrian Academy of Sciences, Schmiedlstrasse 6, A-8042, Graz, Austria\\
$^{8}$Institut d'Astrophysique et de G\'eophysique, Universit\'e de Li\`ege, Quartier Agora (B5c), All\'ee du 6 ao\^ut 19c, 4000 Sart Tilman,\\ Li\`ege, Belgium\\
$^{9}$Instituut voor Sterrenkunde, KU Leuven, Celestijnenlaan 200D, 3001 Leuven, Belgium\\
$^{10}$Dunlap Institute for Astronomy and Astrophysics, University of Toronto, 50 St George Street, Toronto, ON M5S 3H4, Canada\\
}

\date{Accepted XXX. Received YYY; in original form ZZZ}
\pubyear{2017}

\begin{document}
\label{firstpage}
\pagerange{\pageref{firstpage}--\pageref{lastpage}}
\maketitle

\begin{abstract}
In the context of the high resolution, high signal-to-noise ratio, high
sensitivity, spectropolarimetric survey BritePol, which complements observations
by the BRITE constellation of nanosatellites for asteroseismology, we are looking for and measuring
the magnetic field of all stars brighter than V=4. In this paper, we present
circularly polarised spectra obtained with HarpsPol at ESO in La Silla (Chile)
and ESPaDOnS at CFHT (Hawaii) for 3 hot evolved stars: $\iota$\,Car, HR\,3890,
and $\epsilon$\,CMa. We detected a magnetic field in all 3 stars. Each star has
been observed several times to confirm the magnetic detections and check for
variability. The stellar parameters of the 3 objects were determined and their
evolutionary status was ascertained employing evolution models computed with the Geneva code.
$\epsilon$\,CMa was already known and is confirmed to be magnetic, but our
modeling indicates that it is located near the end of the main sequence, i.e.
it is still in a core hydrogen burning phase. $\iota$\,Car and HR\,3890 are the first 
discoveries of magnetic hot supergiants located well after the end of the main
sequence on the HR diagram. These stars are probably the descendants of main
sequence  magnetic massive stars. Their current field strength (a few G) is
compatible with magnetic flux conservation during stellar evolution. These results provide observational constraints for the development of future evolutionary models of hot stars including a fossil magnetic field.
\end{abstract}



\section{Introduction}\label{intro}

The evolution of OB stars on the Hertzsprung-Russell (HR) diagram depends on their initial mass,
metallicity, rotation, and mass loss. The two latter parameters are affected by
the presence and evolution of a stellar magnetic field. Stellar evolution can
also be influenced by the presence of a companion \citep{langer2012}. Because of the many
parameters and effects at play in the evolution of OB stars, it remains
difficult to establish clear evolutionary sequences among observed spectral
types. Stellar evolution models, however, help to trace how these hot stars
evolve. In particular, \cite{groh2014} and \cite{martins2017} showed that, for a
star with an initial mass above 60 M$_\odot$, the supergiant classification
(luminosity class I) is already attributed to stars that are still on the main
sequence (MS). 

However, in current stellar evolution models, the role of the
fossil magnetic field observed in $\sim$10\% of hot stars
\citep{grunhutneiner2015,grunhut2017} is generally not taken into account. Attempts have been made to include a Taylor-Spruit dynamo field in stellar evolution codes \citep[e.g.][]{maeder2003,heger2005}, however it has been shown that such dynamos likely do not exist in MS hot stars \citep{zahn2007, neiner2015}. The effect of a stable dipolar field has also been recently investigated in stellar models, but mainly as a surface effect impacting the wind \citep{petit2017,georgy2017} or for its impact on a given structure \citep{duez2010}.

The fossil fields of hot MS stars are expected to be highly influential in the
context of stellar structure and evolution \citep[e.g.][]{moss1984,maeder2014}.
The basic consequences of magnetic fields for stellar evolution fall into two
general categories: (i) interaction of interior fields with interior fluid
motions, impacting the internal rotational profile, angular momentum, and
chemical transport \citep[e.g.][]{mestel1999,mathiszahn2005,sundqvist2013}; and (ii)
interaction of surface fields with the stellar wind, leading to magnetic braking
of surface layers and reduction of the surface mass-loss rate
\citep{uddoula2002,uddoula2008,uddoula2009,meynet2011}. For example, studies of
magnetic Herbig Ae/Be have shown that their magnetic field
brakes their rotation rate during the early phases of their lives \citep{alecian2013}, and evidence
for spindown on the MS has also been found in hotter stars
\citep[e.g.][]{townsend2010}. Recent studies also suggest, and in some cases
demand, that magnetic fields have direct and ubiquitous consequences for
evolution. For example, in order to explain the post-MS gap of blue supergiants,
\cite{petermann2015} have proposed that these objects evolve
from magnetic MS stars. In addition, \cite{maeder2014} have examined the
role of strong, organized fields in the cores of red supergiants, with
implications for the general spin rates of (magnetic) white dwarfs and pulsars.
Ultimately, these effects lead to important modification of stellar evolutionary
pathways and stellar feedback effects, such as mechanical energy deposition in
the interstellar medium and supernova explosions \citep{heger2005}, and hence the properties
of stellar remnants and potentially the structure and chemistry of the local
Galactic environment.

As magnetic fields influence stellar evolution, so are magnetic fields expected
to transform in response to changes in the structure of the stars in which they
are embedded. In particular, as hot stars age, their radius expands dramatically. The
evolution of surface magnetic fields of hot stars during the MS has been
investigated by \cite{bagnulo2006}, \cite{landstreet2007,landstreet2008}, and
\cite{fossati2016}. These studies provide convincing evidence that the
strengths of surface magnetic fields decrease systematically during the MS, in
response to stellar expansion (the stellar radius typically expands by a factor of $\sim$3 between the zero age MS and the terminal age MS), and possibly due to Ohmic decay and other (currently
unknown) mechanisms. Because hot giants and supergiants retain the radiative
envelopes they had on the MS, evolved OBA stars provide a capability to directly
extend the existing studies of MS objects \citep[such as MiMeS,][]{grunhut2017}
to more advanced evolutionary phases and a much greater range of stellar
structural changes. Considering that about 10\% of all hot stars host a fossil
magnetic field on the MS \citep{grunhutneiner2015,fossati2015bob,grunhut2017}, it is expected
that a similar fraction of hot supergiants could also show magnetic fields of
fossil origin at their surface.

In addition, as hot stars age more and more, convective regions appear in their
radiative envelope. Local dynamos probably develop in these convective zones,
providing an opportunity to also study the unique interactions between the
post-MS dynamo and the pre-existing fossil field. In particular, it is expected
that the dynamo-fossil interaction could enhance the local dynamo fields and
modify the configuration of the global fossil field
\citep{featherstone2009,auriere2008}. Extensive surveys of magnetic fields in
cool (F, G, K) giants and supergiants (the even older evolutionary descendants
of OBA MS stars) have already shown that many of these cool stars exhibit
magnetic fields powered by dynamos \citep[e.g.][]{grunhut2010,auriere2015}.
Although a small population of red giants still show evidence of fossil fields
surviving from the MS \citep[e.g.][]{auriere2008}, the growth of a fully
convective envelope, which replaces the radiative envelope, appears to
effectively erase evidence of their earlier magnetic characteristics at the
surface, even though the fossil field probably still exists inside the star.

Understanding the evolution of the fossil magnetic fields observed in hot MS 
stars is thus important to understand the evolution of these stars. Finding and
understanding magnetic hot supergiants is therefore crucial. However, only two
claimed OB supergiants have been found to host a magnetic field so far. The O9.5
supergiant $\zeta$\,Ori\,Aa hosts a dipolar magnetic field of $\sim$140 G at the
pole \citep{bouret2008,blazere2015}. However, its radius is only $\sim$20 R$_\odot$ \citep{hummel2013}; therefore it appears to be a very young
supergiant. According to the Bonn stellar evolution models, this star is indeed only at about the middle of its MS lifetime \citep[see Fig.~2 of][]{fossati2015}. In the same way, the B1.5 star $\epsilon$\,CMa was found to be
magnetic with a polar field strength of at least 13 G, but
its evolutionary status is unclear: either it is still in its MS phase or
already a post-MS object \citep{fossati2015}. Attempts to detect magnetic fields
in other hot supergiants led to no detection \citep[e.g.][]{grunhut2010,shultz2014}, likely due to
the insufficient precision of the spectropolarimetric measurements compared to the
expected field strength. As of today, there are thus no clearly evolved magnetic
hot stars known.

In this paper, we present new observations of $\epsilon$\,CMa and of two other
hot supergiant stars: $\iota$\,Car and HR\,3890 (Sect.~\ref{obs}). We report 
the determination of their stellar parameters and evolutionary status
(Sect.~\ref{param}), and the detection of their magnetic field
(Sect.~\ref{results}). We then discuss the link between magnetic hot supergiants
and MS magnetic massive stars in the framework of the fossil and dynamo
field scenarios (Sect.~\ref{discus}). 

\section{Observations}\label{obs}

\subsection{BritePol, the BRITE spectropolarimetric survey}

The BRITE (BRIght Target Explorer) constellation of nano-satellites focuses on
the monitoring of  stars with V$\le$4, with high-precision, high-cadence
photometry, in order to perform asteroseismology and to study stellar variability due to stellar and planetary companions, tidal interactions, and rotation  \citep{weiss2014}. The
BRITE sample consists of apparently bright stars, therefore it is dominated by hot stars at all evolutionary stages and evolved
cooler stars (cool giants and AGB stars), which are the most
intrinsically luminous stars.

In this framework, we are performing a spectropolarimetric survey of all targets
with V$\le$4, i.e. about 600 stars, with the goal of discovering new bright
magnetic stars and thus providing prime targets for BRITE seismic studies.
Discovering new bright magnetic stars is also important to provide targets for
multi-technique studies, e.g. including both spectropolarimetric and
interferometric observations.

Spectropolarimetric observations of each of the $\sim$600 targets
(V$\le$4), that are not yet available from archives, have been gathered with
three high-resolution spectropolarimeters: Narval at the
T\'elescope Bernard Lyot (TBL) in France, ESPaDOnS at the Canada-France-Hawaii
Telescope (CFHT) in Hawaii, and HarpsPol on the ESO 3.6-m telescope in La Silla.

In this paper, we report the detection of magnetic fields in 3 hot supergiant stars obtained with
HarpsPol and ESPaDOnS.

\subsection{Spectropolarimetric observations}

The spectropolarimetric observations presented here have been acquired with ESPaDOnS at CFHT
(Hawaii) and HarpsPol at the ESO 3.6-m telescope (La Silla, Chile). The log of
observations is available in Table~\ref{tableobs}.

ESPaDOnS covers a wavelength range from about 375 to 1050 nm, with a resolving
power of $\sim$68000, spread over 40 echelle orders projected onto a single detector. HarpsPol covers a shorter
wavelength range from about 380 to 690 nm on two detectors and 71 echelle
orders, but with a higher resolving power of $\sim$110000.

\begin{table}
\caption{Journal of observations indicating the name of the stars, the instrument used for the spectropolarimetric measurements (H=HarpsPol, E=ESPaDOnS), the Heliocentric Julian Date at the middle of the observations (mid-HJD - 2450000), the exposure time in seconds, and the average signal-to-noise ratio of a spectropolarimetric sequence per CCD pixel at $\sim$500 nm.}
\begin{tabular}{ll@{\,\,}l@{\,\,}ll@{\,\,}r}
\hline
Star & Inst. & Date & mid-HJD & T$_{\rm exp}$ & S/N \\
     &       &      & -2450000 &  &  \\
\hline
$\iota$\,Car	  & H & Nov 9, 2014 & 6970.8666 & 4$\times$34    & 420 \\
$\iota$\,Car	  & H &Nov 10, 2014 & 6971.8472 & 5$\times$4$\times$55  & 510 \\
$\iota$\,Car	  & H & Mar 2, 2015 & 7083.6199 & 4$\times$55    & 500 \\
$\iota$\,Car	  & H & Mar 3, 2015 & 7084.5692 & 3$\times$4$\times$75  & 560 \\
$\iota$\,Car	  & H & Mar 9, 2015 & 7090.6508 & 3$\times$4$\times$75  & 700 \\
HR\,3890  	  & H & Mar 9, 2015 & 7090.6638 & 4$\times$119   & 650 \\
HR\,3890  	  & H &May 13, 2015 & 7155.5660 & 4$\times$125   & 725 \\
$\epsilon$\,CMa	  & E & Feb 8, 2014 & 6696.8619 & 4$\times$7	 & 775 \\
$\epsilon$\,CMa	  & E & Nov 9, 2014 & 6971.1382 & 9$\times$4$\times$7   & 1270\\
$\epsilon$\,CMa	  & E &Dec 22, 2014 & 7014.0850 & 9$\times$4$\times$7   & 1080 \\
$\epsilon$\,CMa	  & E & Jan 9, 2015 & 7032.0893 & 9$\times$4$\times$7   & 1010 \\
\hline
\end{tabular}
\label{tableobs}
\end{table}

We observed the targets in circular polarisation mode. Each observation consists
of four sub-exposures taken with the polarimeter in various configurations. The four
sub-exposures are constructively combined to obtain the Stokes V spectrum and 
destructively combined to produce a null polarisation profile (N) to check for
pollution by, e.g., instrumental effects, variable observing conditions, or
physical phenomena unrelated to magnetism such as pulsations. In
addition, the intensity (Stokes I) spectrum is extracted. Finally, successive sequences can be acquired and co-added to increase the signal-to-noise ratio (S/N) of a magnetic measurement.

\begin{table*}
\caption{Parameters for the three stars $\iota$\,Car, HR\,3890, and $\epsilon$\,CMa. Columns report the star name, spectral type, effective temperature, gravity, luminosity, mass at the ZAMS, rotation rate at the ZAMS, radius at the ZAMS, current mass, current radius, and age. For each star, the results from the two most extreme evolutionary models reproducing simultaneously T$_{\rm eff}$, $\log(g)$, and $\log(L/L_\odot)$ are shown, to illustrate the range of possible initial conditions. }
\begin{tabular}{@{\,}l@{\,\,}lllll@{\,\,}l@{\,\,}l@{\,\,}lll@{\,}}
\hline
Star & SpT & T$_{\rm eff}$ & $\log(g)$ & $\log(L/L_\odot$) & M$_{\rm ZAMS}$ & \multirow{2}{*}{$\frac{\Omega_{\rm ZAMS}}{\Omega_{\rm crit}}$} & R$_{\rm ZAMS}$ & M & R & age\\
 & & K & cm~s$^{-2}$ & & M$_\odot$ & & R$_\odot$ & M$_\odot$ & R$_\odot$ & Myr\\
\hline
$\iota$\,Car & A7Ib$^1$ & $7500\pm100$ & $1.85\pm0.10$ & $4.53\pm0.37^2$ & 7.0 & 0.30 & 3.00 & 6.90 & 46.4-50.1 & 56.35-56.40\\
& & & & & 11.0 & 0.10 & 3.82 & 10.96 & 69.1-72.7 & 18.88\\
HR\,3890 & A7Ib & $7500\pm100$ & $1.4\pm0.1$ & -- & 11.0 & 0.9 & 4.32 & 10.88 & 96.6-100.0 & 23.42-23.44\\
& & & & & 15.0 & 0.00 & 4.48 & 14.70 & 134.6-141.3 & 11.52-11.53\\
$\epsilon$\,CMa & B1.5II & $22500\pm300^3$ & $3.4\pm0.08^3$ & $4.35\pm0.05^3$ & 12.0 & 0.30 & 4.00 & 11.92 & 10.4-10.5 & 17.62-17.88\\
& & & & & 12.0 & 0.80 & 4.31 & 11.91 & 10.4 & 18.86-18.96\\
\hline
\end{tabular}\\
Notes: taken from $^1$\cite{GrGa89}, $^2$\cite{Boyarchuk1984a}, and  $^3$\cite{fossati2015}.
\label{tableparam}
\end{table*}

The usual calibrations (bias, flat-field, and ThAr frames) have been obtained each night 
and applied to the data. The ESPaDOnS data reduction was performed using {\sc
Libre-Esprit} \citep{donati1999} and {\sc Upena}, a dedicated software pipeline
available at CFHT. The Stokes I spectra were normalized to the continuum
level, order by order, using {\sc IRAF}\footnote{IRAF is distributed by the
National Optical Astronomy Observatory, which is operated by the Association of
Universities for Research in Astronomy (AURA) under a cooperative agreement with
the National Science Foundation.}, and the same normalization function was applied to the
Stokes V and N spectra. The HarpsPol reduction was performed with a modified
version of the {\sc REDUCE} package \citep{piskunov2002,makaganiuk2011}.
Automatic normalization was first performed with {\sc REDUCE} and an additional
normalization function was fitted with {\sc IRAF} to improve the final normalization.

Finally, we applied the Least Squares Deconvolution (LSD) technique 
\citep{donati1997} to produce the mean LSD Stokes I, Stokes V, and N
profiles of each magnetic measurement. Consecutive sequences of observations of the same star were then co-added
to produce one single magnetic measurement. 

LSD requires a list of all lines present in the spectrum, including their  wavelength, depth,
and Land\'e factor. Such a line mask was produced for each of the three targets. We first
extracted lists of lines from the VALD3 atomic database \citep{piskunov1995,
kupka1999} for the appropriate temperature and gravity of each star (determined in Sect.~\ref{param} below, also see Table~\ref{tableparam}). We restricted the lists to lines with a depth larger than 0.01. We then removed from each mask all
lines that are not present in the observed intensity spectra, hydrogen lines because  
their profile is different from those of metal lines, and lines blended with either H lines, interstellar lines, or telluric lines. In
addition, in the case of the mask for the hottest star $\epsilon$\,CMa,  we
removed the He lines with broad wings \citep[see][and Wade et al. in prep.]{fossati2015}. Finally, the
depth of each line in the LSD masks was adjusted so as to fit the observed line
depth. This method is described in more details in \cite{grunhut2017}.

\section{Stellar parameters}\label{param}

\subsection{Determination of the atmospheric parameters}\label{atmos}

The star $\iota$\,Car was classified as A7\,Ib by \citet{GrGa89}. HR\,3890 is the supergiant primary of a visual binary, with the 3.2\,mag fainter B3-4\,IV secondary \citep[HR\,3891,][]{MaNi86} located only 5$\arcsec$ away. The companion, however, is sufficiently far away not to be recorded in the HarpsPol spectra presented here, since Harps's fibers have a diameter of 1$\arcsec$. The spectra of $\iota$\,Car and HR\,3890 are almost identical, except for some metal lines and the Balmer wings that indicate a slightly higher luminosity for HR\,3890. The H$\alpha$ profile is symmetric and in absorption. We therefore assign a spectral type of A7\,Ib to HR\,3890 as well, similar to the earlier A8\,Ib classification by \citet{HoCo75}. $\epsilon$\,CMa is one of the two primary standards defining the B1.5\,II  class in the MK system \citep{GrCo09}. Both the spectral types and the luminosity classification provide guidance for the starting points of the quantitative analysis.

The determination of the atmospheric parameters of $\iota$\,Car and HR\,3890 is based on the hybrid non-LTE (local thermodynamic equilibrium) approach described by \citet{Prz06} and \citet{FiPr12} for the analysis of BA-type supergiants.
In brief, the modelling is based on plane-parallel, hydrostatic, chemically homogeneous and line-blanketed model atmospheres that were computed with the code {\sc Atlas9} \citep{Kurucz93} under the assumption of LTE. Sphericity may be considered an issue for the extended atmospheres of supergiants.
However, these two stars are near the lower luminosity boundary of the supergiant class. The atmospheres have an extension of only 1--2\% of the stellar radius, therefore deviations from plane-parallel geometry can be considered negligible. The onset  of convection in stellar atmospheres occurs among the mid/late A-type stars.
We therefore considered convective energy transport using the standard B\"ohm-Vitense mixing-length theory \citep{boehm1958} employed in the {\sc Atlas9} code, adopting a ratio of mixing length to pressure scale height of 1.5. The fraction of the total energy flux transported by convection  remains below 1\% in both cases.

In a second step, non-LTE level populations and synthetic spectra were calculated with recent versions of the codes {\sc Detail} and {\sc Surface} \citep{Gid81,BuGi85}. Here, essential extensions to the  work of \citet{Prz06} consist of 
i) the implementation of additional bound-free and free-free opacities appropriate for cooler temperatures (most important H$^-$,  plus metals with low first ionisation threshold), 
ii) consideration of collisions with neutral hydrogen atoms \citep[using the formula of][]{StHo84},
iii) inclusion of self-broadening for the Balmer lines \citep{Bar00a}, and  
iv) Van-der-Waals broadening using coefficients from \citet{Bar00b}, the collection of Kurucz\footnote{\tt http://kurucz.harvard.edu/linelists.html} or approximations \citep[see, e.g.,][]{Cas05}.

We focused on simultaneous fits to the Stark-broadened hydrogen Balmer lines and the Mg\,{\sc i/ii} ionization equilibrium in order to constrain the effective temperature $T_{\rm eff}$ and surface gravity $\log g$. Model atoms as described by \citet{Prz01} and \citet{PrBu04} were employed for the two species, with the above mentioned modifications. The derived atmospheric parameter values are summarized in Table~\ref{tableparam}. Our $T_{\rm eff}$ value for $\iota$\,Car is in good agreement with previous determinations by \citet{LuLa85,LuLa92}, \citet{Smi06} and \citet{TaBa16}, while our surface gravity value is higher than in most of these works ($\log g = 0.9$ to 1.6), and significantly  lower than the  $\log g = 2.3$ derived by \citet{Smi06}. In the case of HR\,3890, atmospheric  parameters have so far been reported only by \citet{Sam88},  $T_{\rm eff} = 7600 \pm 350 $K and $\log g =1.1 \pm 0.3$, which agree  with our results within the uncertainties.

The atmospheric parameter determination for $\epsilon$\,CMa was described in detail by \cite{fossati2015} and these parameters are reported in  Table~\ref{tableparam}. The same model codes as employed here for the A-type supergiants were used for $\epsilon$\,CMa by \cite{fossati2015}, following the analysis strategy and adopting model atoms as summarized by \citet{NiPr12}.

\subsection{Evolutionary status}

\begin{figure*}
\includegraphics[width=.48\textwidth]{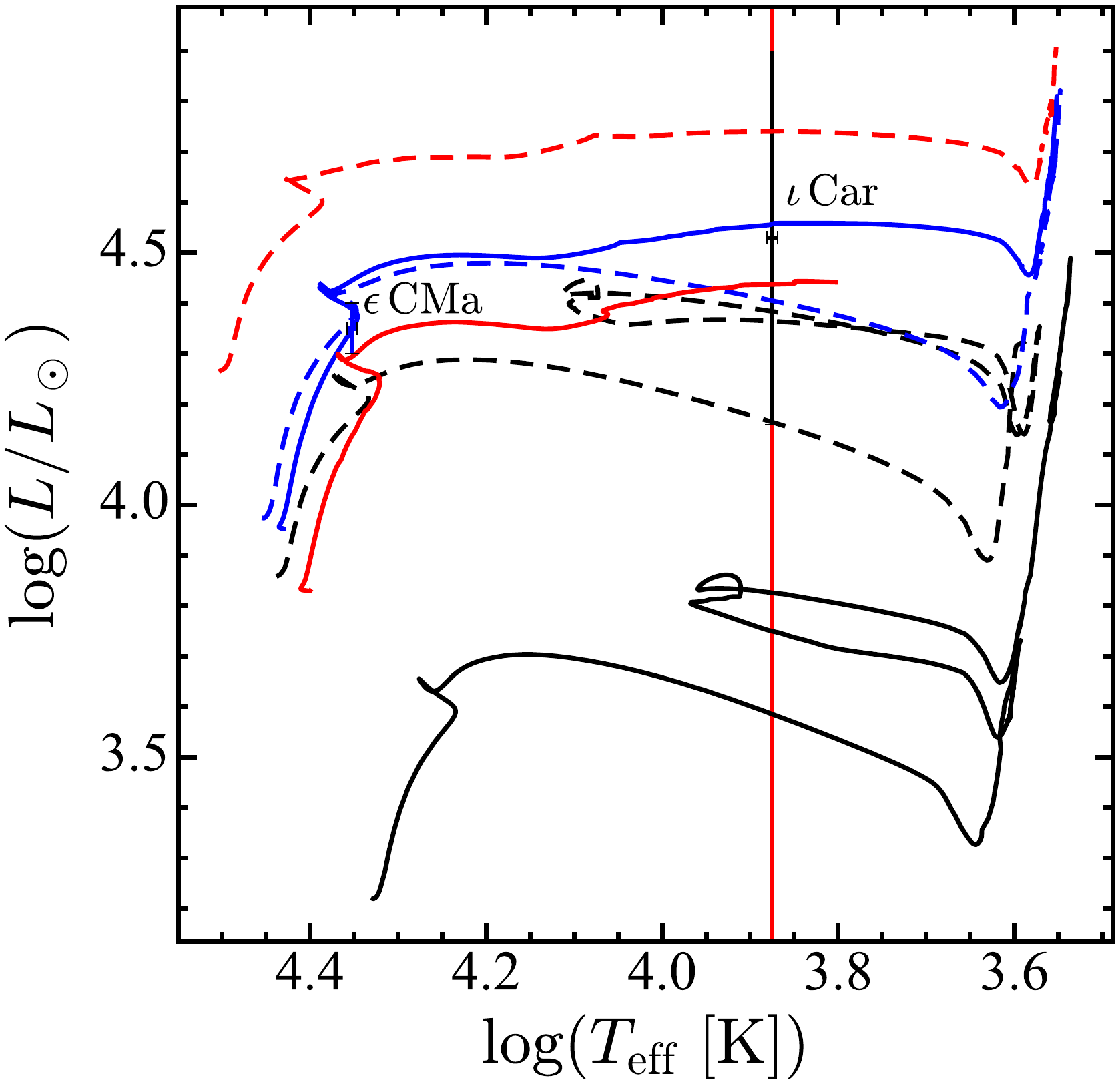}
\hfill
\includegraphics[width=.48\textwidth]{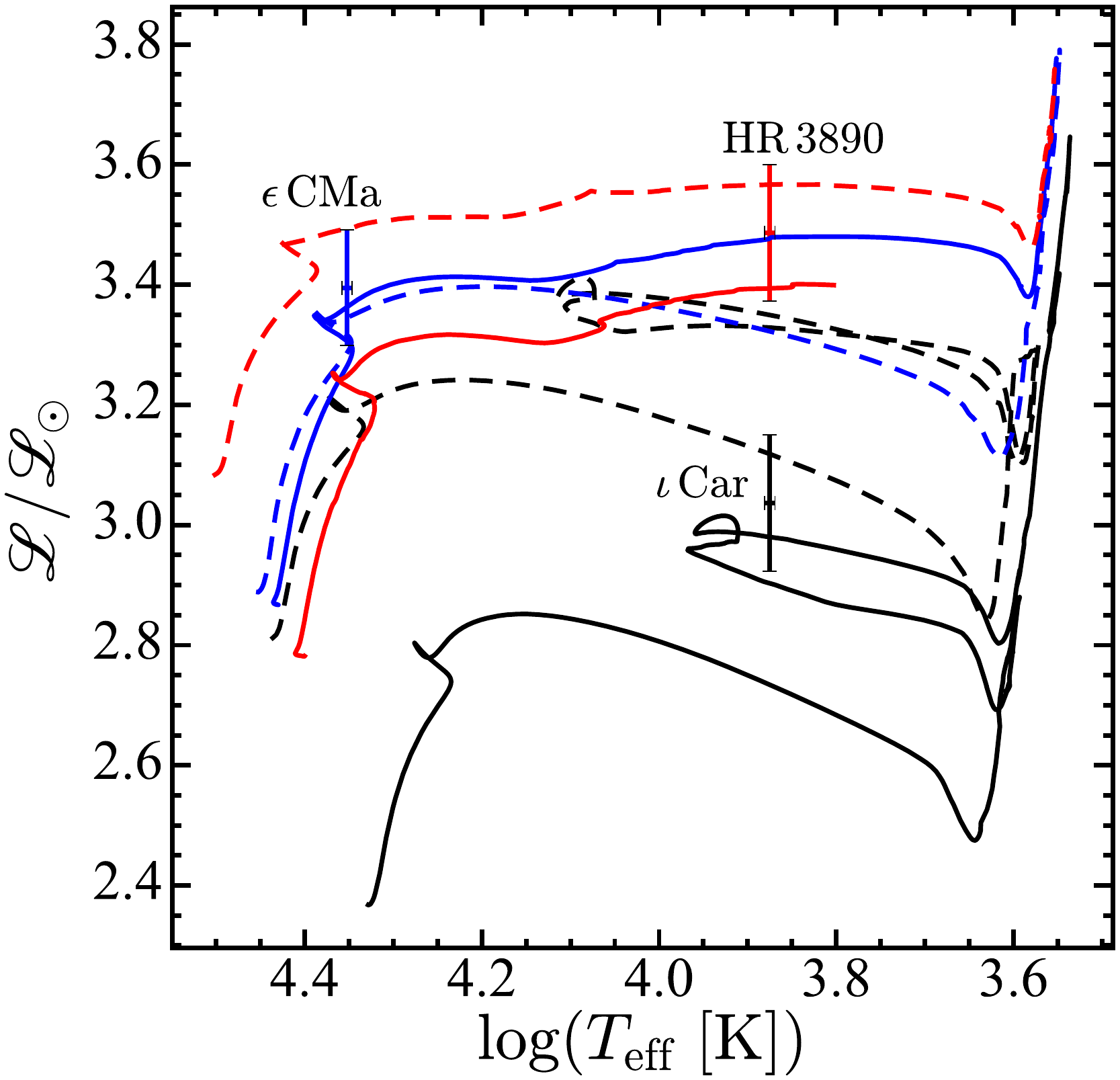}
\caption{\textit{Left panel:} HR diagram of our models compared to the observed
position of $\iota$ Car and $\epsilon$ CMa. The black tracks correspond to the
two most extreme models fitting the position of $\iota$\,Car:
7 M$_\odot$ with $\Omega_{\rm ZAMS}/\Omega_{\rm crit} = 0.30$ (solid)
and 11 M$_\odot$ with $\Omega_{\rm ZAMS}/\Omega_{\rm crit} = 0.10$ (dashed). The blue tracks correspond to
the two most extreme models fitting the position of $\epsilon$\,CMa:
12 M$_\odot$ with $\Omega_{\rm ZAMS}/\Omega_{\rm crit} = 0.8$ (solid)
and 12 M$_\odot$ with $\Omega_{\rm ZAMS}/\Omega_{\rm crit} = 0.3$
(dashed). The red tracks corresponds to the two most extreme models fitting the
position of HR\,3890: 11 M$_\odot$ with
$\Omega_{\rm ZAMS}/\Omega_{\rm crit} = 0.9$ (solid) and non-rotating
15 M$_\odot$ (dashed). As we have found no measurement of the luminosity
of HR\,3890 in the literature, we only show its effective temperature with the thin vertical red line. Note, however, that the spectrum of HR\,3890 indicates a slightly higher luminosity than for $\iota$\,Car. \textit{Right
panel:} tracks of our models in the spectroscopic HR diagram \citep{Langer2014a}, where $\mathcal{L} = T_{\rm eff}^4/g$. The observational errors on $\mathcal{L}$ are computed as $\Delta\mathcal{L} = \sqrt{\left(\frac{4\Delta T_{\rm eff}}{T_{\rm eff}}\right)^2+\Delta\log(g)^2}$. The
colors and line styles are the same as in the left panel.}
\label{Fig_CompaObs}
\end{figure*}

Determining the evolutionary status and stellar parameters, such as the initial
mass, current age, etc., for post-MS phases of massive stars on the basis of
stellar evolution models is not a straightforward task, and is extremely
dependent on the stellar evolution code used to build the models, due to various
ways of implementing physical processes (convection, rotation, overshooting,...) inside such
codes.

In this paper, we used different sets of models computed with the Geneva stellar
evolution code \citep{Eggenberger2008a} at solar metallicity and including the
effects of rotation \citep{Ekstrom2012a,Georgy2013a}. To refine our
determination of the evolutionary status, we have also computed several
additional models, using exactly the same code and physical
ingredients\footnote{These models, however, were computed only up to the base of
the red supergiant branch, which is enough for the purpose of this paper.}. We stress that none of the models presented in this paper include a magnetic field, since evolutionary models computed including the physical influence of a fossil field in the radiative envelope of hot stars do not exist yet (see Sect.~\ref{intro}).

For
each star, we fit simultaneously the available observational data. The effective
temperature and surface gravity were determined from our spectra (see
Sect.~\ref{atmos}). We also accounted for the luminosity when
this quantity was available from the literature, i.e. for $\iota$\,Car and $\epsilon$\,CMa.

The results for each individual star are discussed below. The last six columns of Table~\ref{tableparam} 
give the initial mass, the initial rotation rate, the initial radius, the current mass, the current radius, and the permitted 
age range. For each star, the results of the two most extreme models (smaller and higher initial mass)
are shown. In Fig.~\ref{Fig_CompaObs}, we show how the tracks compare to the
observations in the classical (left) and spectroscopic (right) HR diagram.

\subsubsection{$\iota$ Car}

$\iota$\,Car has an effective temperature $T_{\rm eff} = 7500 \pm 100$ K and a
surface gravity of $\log(g) = 1.85 \pm 0.10$ from our determinations. From this
determination, the constraint on the models are quite weak (right panel of
Fig.~\ref{Fig_CompaObs}): from $\sim$7 M$_\odot$ (solid black curve) up
to $\sim$11 M$_\odot$ (dashed black curve). Adding a constraint on the
luminosity \citep[$\log(L) = 4.53\pm0.37$,][]{Boyarchuk1984a} allows us to tighten
the estimate (left panel), limiting the mass to about 11 M$_\odot$. We remark that the luminosity estimate from \citet{Boyarchuk1984a}
was based on a different set of stellar models and does not come from direct
independent measurements.

In all cases, $\iota$\,Car is clearly a post-MS star. Depending on its mass, it
could be on its first crossing of the HR diagram (if its mass is closer to
11 M$_\odot$), or on a blue loop (if its mass is closer to
7 M$_\odot$). Its age is between $\sim$19 and 56 Myr,
depending on its mass. Other quantities such as the current mass and radius can
be found in Table~\ref{tableparam}. On the MS, $\iota$\,Car was a B star. 

\subsubsection{HR\,3890}

In the case of HR\,3890, no distance is available to determine its luminosity.
Therefore, we rely on our estimates of $T_{\rm eff}$ and $\log(g)$ only.
For these two quantities, we have obtained $T_{\rm eff} =
7500\pm100$ K and $\log(g) = 1.4\pm0.1$, which provide good constraints (right panel of Fig.~\ref{Fig_CompaObs}). The star is also clearly a post-MS star. According to the set of models we used, its mass is higher than the maximal mass for which a Cepheid blue loop occurs. This star is therefore on its first (and unique) crossing of the HR diagram. The
range of possible models goes from a rotating 11 M$_\odot$ star with an
initial angular velocity of 90\% of the critical velocity to a non-rotating
15 M$_\odot$ star. The age would be between $\sim$11.5 Myr up to $\sim$23.5 Myr, depending on the initial mass.
The current radius of the star ranges from $\sim$96 to $\sim$141 R$_\odot$. On the MS, HR\,3890 was a B star. 

\subsubsection{$\epsilon$ CMa}

For $\epsilon$ CMa, we used the stellar parameters from \citet{fossati2015}.
They obtained $T_{\rm eff} = 22500 \pm 300$ K, $\log(g) =
3.40\pm 0.08$, and $\log(L/L_\odot) = 4.35 \pm 0.05$. From our evolutionary models,
we obtain that the initial mass of the star should be very close to
12 M$_\odot$, with an initial angular velocity in the range of 30\% to
80\% of the critical velocity. The only solution simultaneously fitting the effective
temperature, the luminosity and the surface gravity is a model near the end of
the MS (see both panels of Fig.~\ref{Fig_CompaObs}). The age of the
star is between $\sim$17.5 and $\sim$19 Myr, and the current radius is
around 10.5 R$_\odot$.

\begin{table}
\caption{Results of the magnetic field measurements for the 3 stars $\iota$\,Car, HR\,3890 and $\epsilon$\,CMa. The first four columns provide the name of the star, the longitudinal field ($B_l$) and null ($N_l$) measurements in Gauss, with their error bars, and detection status (DD = definite detection, ND = no detection), for each magnetic measurement. The last two columns indicate the inferred magnetic polar field strength in Gauss, as it currently is ($B_p$) assuming a dipole field and as it was on the MS ($B_{\rm MS}$) assuming magnetic flux conservation, for each star.}
\begin{tabular}{@{\,}ll@{\,}ll@{\,}ll@{\,\,}l@{\,\,\,}l@{\,}}
\hline
Star & $B_l$ & $\pm \sigma B_l$ & $N_l$ & $\pm \sigma N_l$ & Detect. & $B_p$ & $B_{\rm MS}$\\
\hline
$\iota$\,Car	 & -0.8 & $\pm$ 0.5 & -0.3 & $\pm$ 0.5 & ND  & 3 & 700-1100 \\
$\iota$\,Car	 & -0.9 & $\pm$ 0.2 & -0.1 & $\pm$ 0.2 & DD  &  & \\
$\iota$\,Car	 & -0.4 & $\pm$ 0.4 & ~0.1 & $\pm$ 0.4 & ND  &  & \\
$\iota$\,Car	 & -0.5 & $\pm$ 0.2 & ~0.1 & $\pm$ 0.2 & ND  &  & \\
$\iota$\,Car	 & -0.3 & $\pm$ 0.2 & -0.1 & $\pm$ 0.2 & ND  &  & \\
HR\,3890  	 & -1.9 & $\pm$ 1.1 & -0.6 & $\pm$ 1.1 & ND  & 6 & 3000-6000\\
HR\,3890  	 & -1.0 & $\pm$ 1.0 & -1.1 & $\pm$ 1.0 & ND  &  & \\
$\epsilon$\,CMa	 & -1.2 & $\pm$ 10.9& -4.5& $\pm$ 10.9 & DD  & 32 & 185-220 \\
$\epsilon$\,CMa	 & -9.1 & $\pm$ 2.1 & 1.6 & $\pm$ 2.1 & DD  &  & \\
$\epsilon$\,CMa	 & -9.6 & $\pm$ 2.5 & 1.6 & $\pm$ 2.5 & DD  &  & \\
$\epsilon$\,CMa	 & -6.3 & $\pm$ 2.9 & 0.2 & $\pm$ 2.9 & DD  &  & \\
\hline
\end{tabular}
\label{tableresults}
\end{table}

\section{Magnetic analysis and results}\label{results}

For each spectropolarimetric measurement, using a center-of-gravity method \citep{rees1979,wade2000}, we calculate the longitudinal field value $B_l$  corresponding to the Zeeman signatures observed in the LSD Stokes V profiles. We do the same for the N profiles. 

In addition, the detection of a magnetic field is evaluated by the False Alarm Probability (FAP) of a signature in the LSD Stokes V profile inside the velocity range defined by the LSD I line width, compared to the mean noise level in the LSD Stokes V profile outside the line. We adopted the convention defined by \cite{donati1997}: if FAP $<$ 0.001\%, the magnetic detection is definite, if 0.001\% $<$ FAP $<$ 0.1\% the detection is marginal, otherwise there is formally no magnetic detection. 

Results for the three stars are shown in Table~\ref{tableresults} and described below.

\subsection{$\iota$\,Car}

$\iota$\,Car has been observed five times with HarpsPol in November 2014 and March 2015. Each measurement consisted of one or more consecutive Stokes V sequences of 4 subexposures of 34 to 75 seconds each, i.e. a total exposure time between 136 and 1100 seconds per magnetic measurement (see Table~\ref{tableobs}). The line mask produced for this star includes 7014 lines. The LSD profiles have a S/N ranging from 2360 to 5300 in Stokes I, and 15000 to 44000 in Stokes V. 

\begin{figure}
\resizebox{\hsize}{!}{\includegraphics[clip]{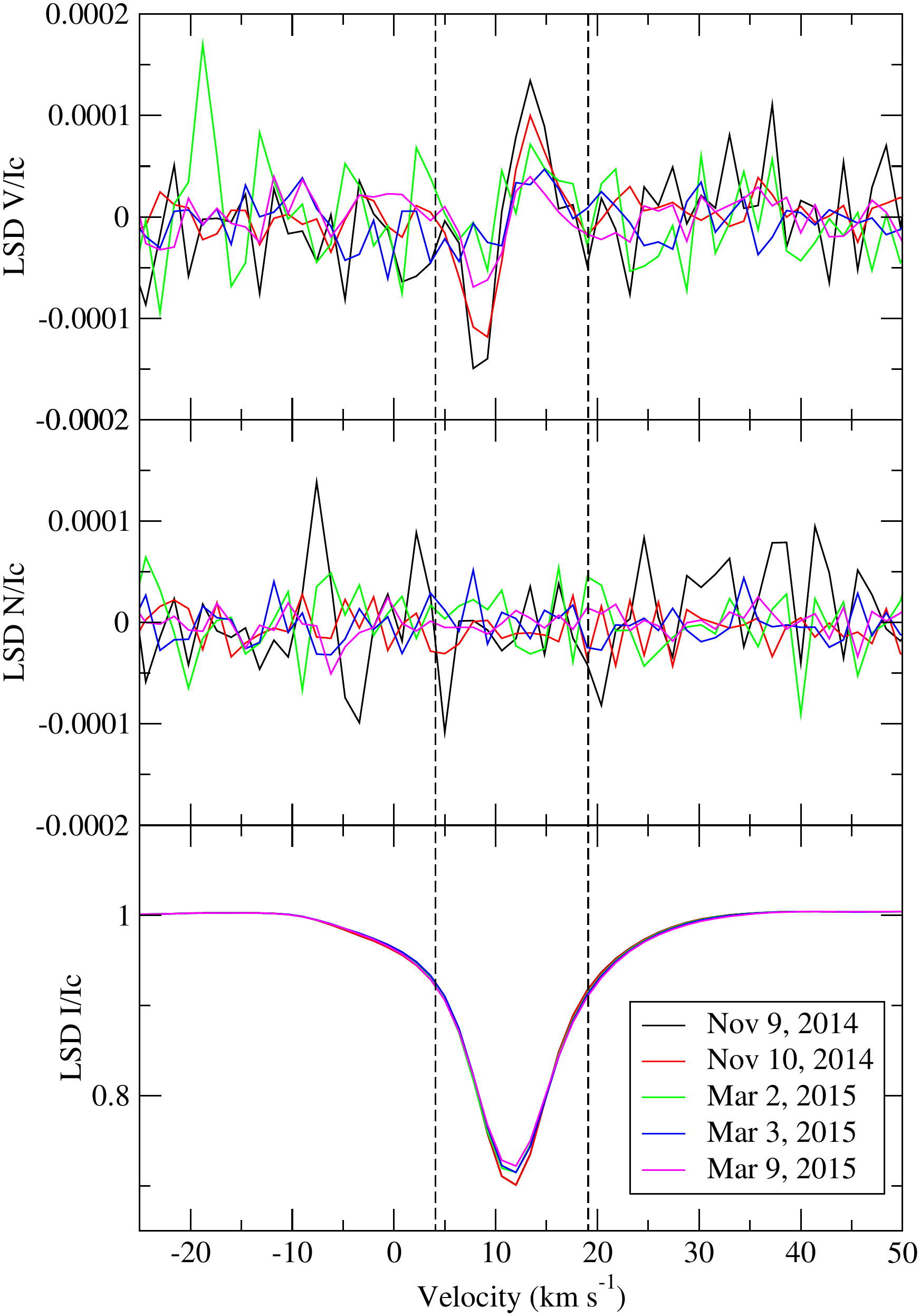}}
\caption[]{LSD Stokes V (top), N (middle) and Stokes I (bottom) profiles for the star $\iota$\,Car, taken on 5 different nights. Vertical dashed lines indicate the integration range employed to calculate the longitudinal field and estimate the FAP.}
\label{iotcar}
\end{figure}

While the magnetic signature is clearly visible in all 5 measurements of $\iota$\,Car (see Fig.~\ref{iotcar}), only the second measurement, taken close to the pole-on phase and with a very high S/N, results formally in a definite detection. For the other measurements, the signature is too weak compared to the noise level in the continuum. Nevertheless, the observed signature, present at the position of the intensity line profile and which varies in time in the way expected from a dipole field, leaves no doubt that $\iota$\,Car is magnetic. The magnetic signature, however, is only visible in the core of the LSD line. Broad wings are often observed in the spectral lines of hot supergiants and could be due to strong macroturbulence \citep[see][]{simondiaz2017}. We also note that small variations are visible in the core of the intensity profile. 

From these Stokes V signatures and the corresponding I profiles, we calculated the longitudinal field values by integrating in a range of $\pm$7.5 km~s$^{-1}$ around the center of the line (at 11.6 km~s$^{-1}$), thus excluding the broad wings. $B_l$ values are reported in Table~\ref{tableresults}. The longitudinal field is systematically negative at the rotational phases at which the data were acquired and below 1 G in strength. Similar measurements performed with the N profiles provide values compatible with 0 (see Table~\ref{tableresults}).

In addition, from the weak variation observed in the Stokes V signatures and in the $B_l$ values for data acquired during a few nights in each of the two periods of observation and 4 months apart between the two periods, we conclude that the rotation period of $\iota$\,Car is probably long, of at least a few months. Follow-up observations of this target should allow an accurate determination of the star's rotation period. 

\subsection{HR\,3890}

\begin{figure}
\resizebox{\hsize}{!}{\includegraphics[clip]{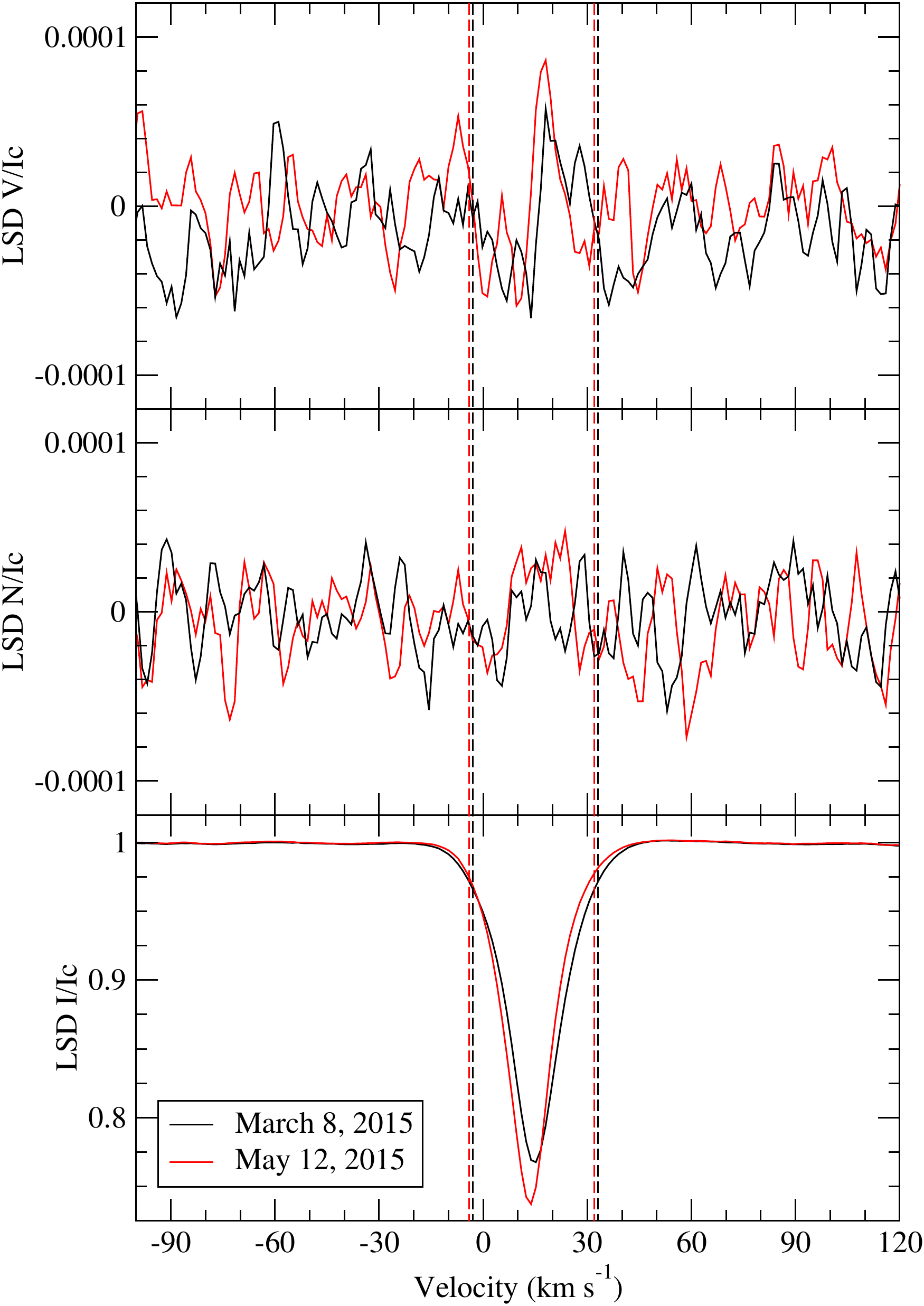}}
\caption[]{Same as Fig.~\ref{iotcar}, but for the two measurements of the star HR\,3890.}
\label{hr3890}
\end{figure}

HR\,3890 has been observed twice with HarpsPol, once in March 2015 and once in May 2015. Each measurement consisted of a single Stokes V sequence of 4 subexposures of either 119 or 125 seconds each, respectively, i.e. a total exposure time of 476 and 500 seconds per magnetic measurement (see Table~\ref{tableobs}). The line mask produced for this star includes 5448 lines. The LSD profiles have a respective S/N of 3060 and 3040 in Stokes I, and 18316 and 20598 in Stokes V. 

A weak but clear signature is visible in the Stokes V profile of both measurements, with a similar shape in spite of the $\sim$2 months difference in observation date and variation in the intensity profile (see Fig.~\ref{hr3890}). 

Following the example of $\iota$\,Car, we excluded the line wings from the calculation of the longitudinal field values. We thus integrated the two Stokes V signatures and the corresponding I profiles in a range of $\pm$18 km~s$^{-1}$ around the center of the line (15 and 14 km~s$^{-1}$, respectively for the two measurements). $B_l$ values are reported in Table~\ref{tableresults}. The field is systematically negative and between -1 and -2 G in strength. Similar measurements ($N_l$) were also performed with the N profiles. Although the second $B_l$ value is compatible with 0 within the error bars and no formal detection is obtained from the FAP analysis for any of the two measurements (see Table~\ref{tableresults}), the fact that Stokes V profiles show a clear and repeatable signature, while N profiles show only noise, provides confidence that the magnetic signature is real. 

\subsection{$\epsilon$\,CMa}

\begin{figure}
\resizebox{\hsize}{!}{\includegraphics[clip]{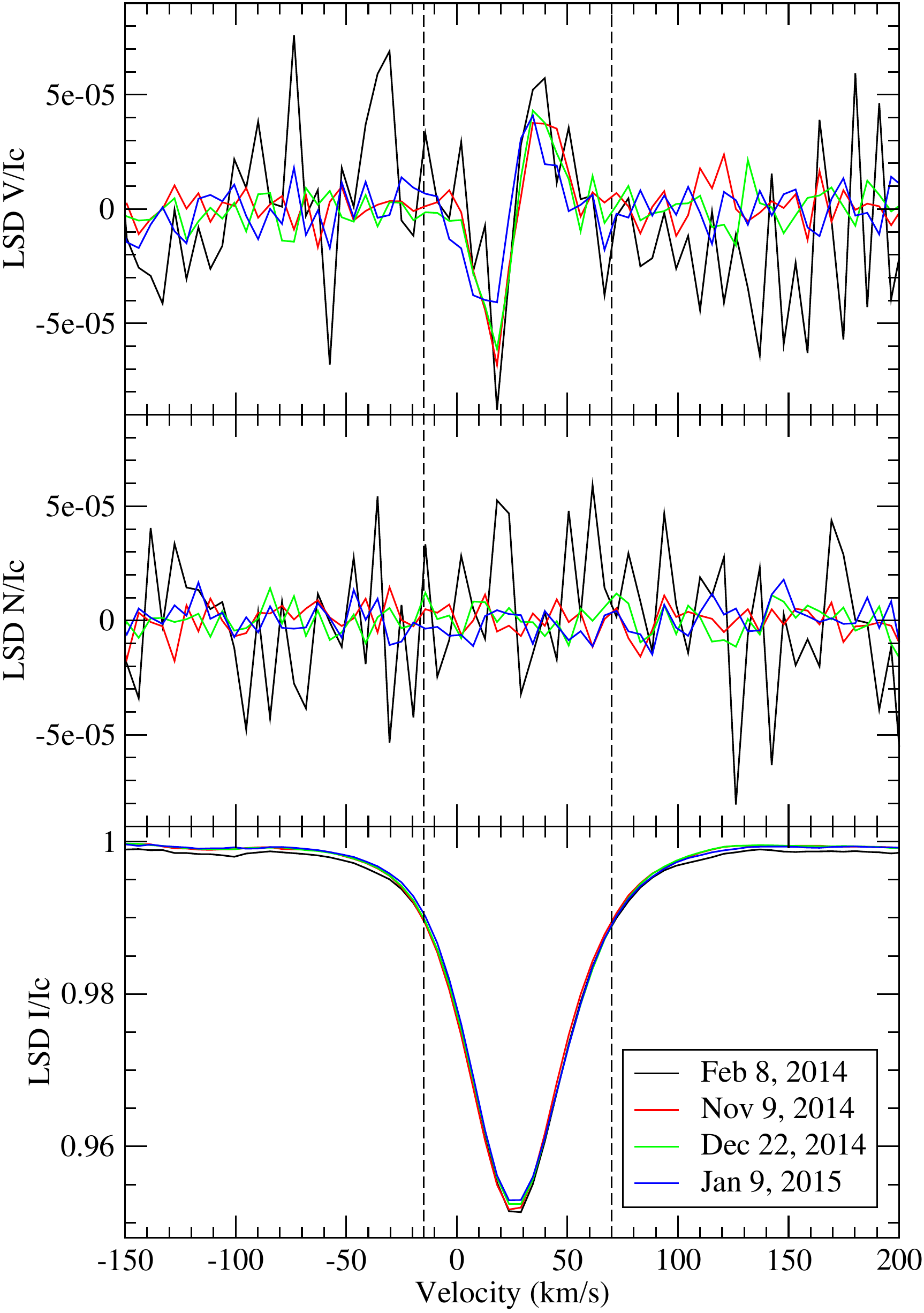}}
\caption[]{Same as Fig.~\ref{iotcar}, but for the four measurements of the star $\epsilon$\,CMa.}
\label{epscma}
\end{figure}

\begin{figure*}
\includegraphics[width=.48\textwidth, angle=0,clip]{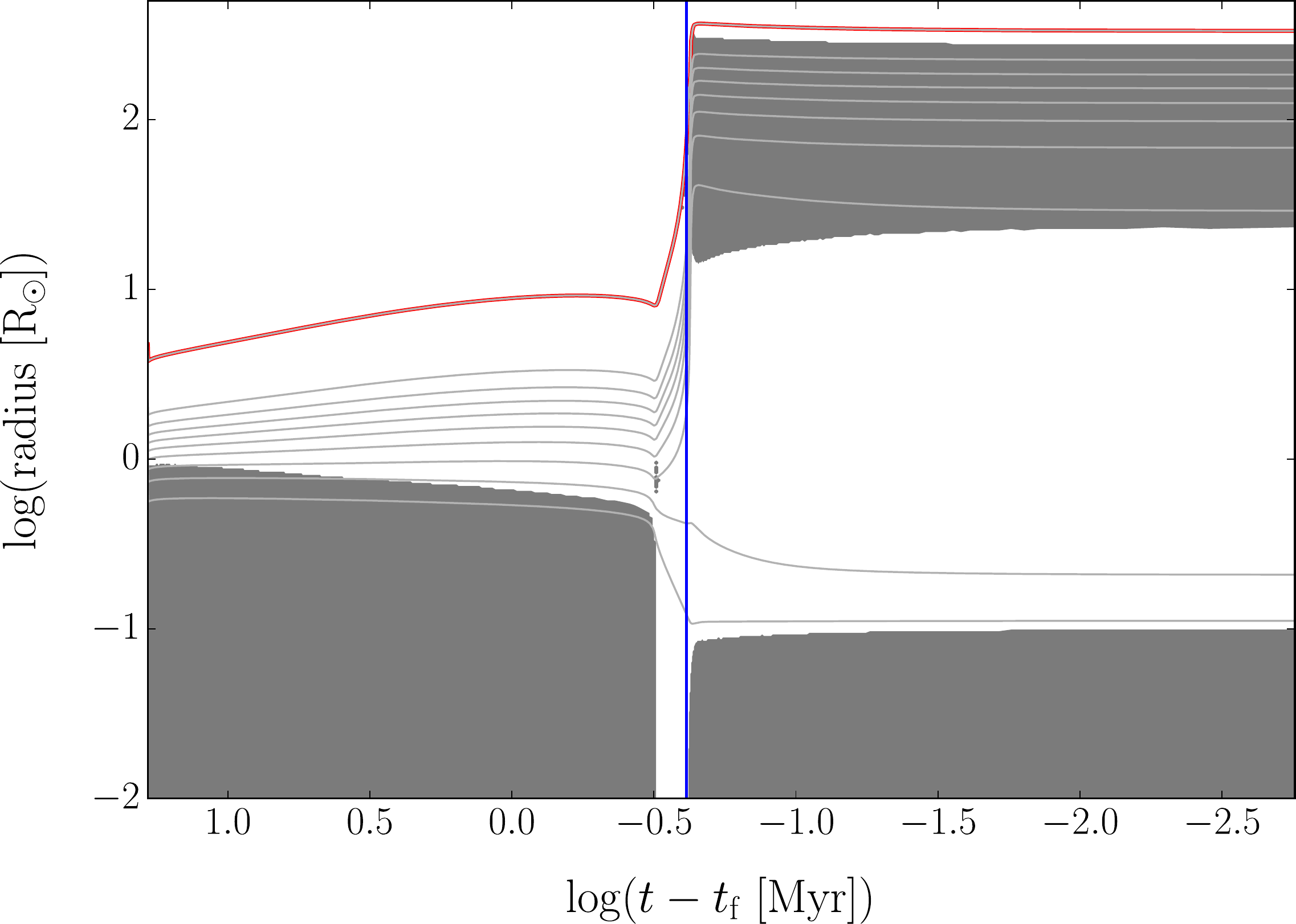}
\includegraphics[width=.48\textwidth, angle=0,clip]{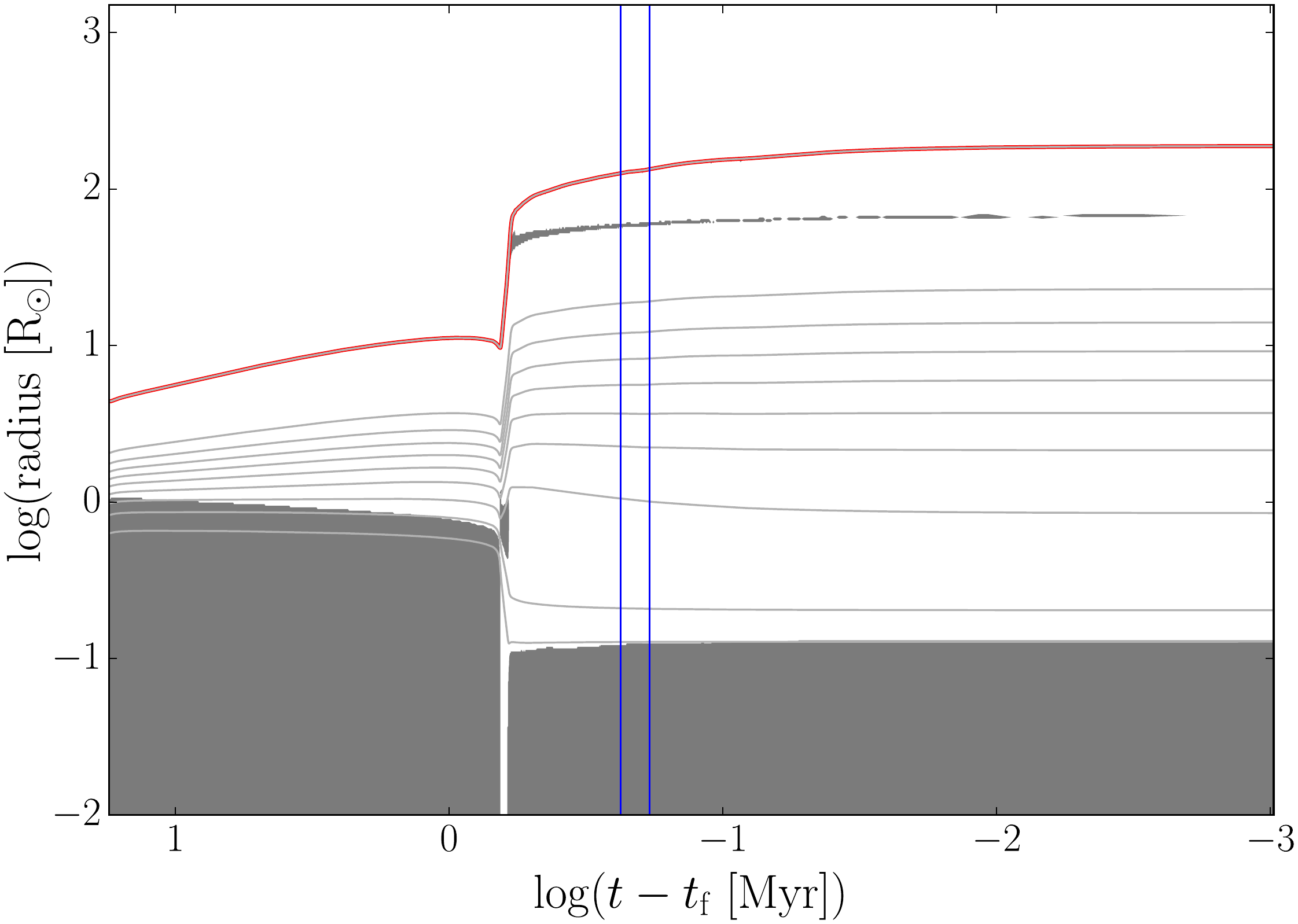}
\caption{Kippenhahn diagrams showing the evolution of the stellar structure of $\iota$\,Car (left) and HR\,3890 (right). Convective zones are indicated in gray, while radiative zones are in white. The red line indicates the surface of the star and the thin gray lines show the mass distribution for 11 regular steps between 0 and the initial mass value of the star. The vertical blue lines show the current position of the star in this diagram, within uncertainties.}
\label{Fig_Kippenhahn}
\end{figure*}

$\epsilon$\,CMa has been observed four times with ESPaDOnS in February, November, December 2014, and January 2015. The first measurement consisted of only one single Stokes V sequence of 4 subexposures of 7 seconds each, i.e. a total exposure time of 28 seconds. The following 3 measurements consisted of 9  consecutive Stokes V sequences of 4 subexposures of 7 seconds each, i.e. a total exposure time of 252 seconds per magnetic measurement (see Table~\ref{tableobs}). The line mask produced for this star includes 1746 lines. The LSD profiles have a S/N from 3989 to 4803 in Stokes I and 23405 to 119309 in Stokes V. 

All four LSD profiles provide a definite detection according to the FAP analysis, with clear magnetic signatures in Stokes V and no signal in N. We integrated the four Stokes V signatures and the corresponding I profiles in a range of 27.5$\pm$42.5 km~s$^{-1}$, i.e. again excluding the broad wings. $B_l$ values are reported in
Table~\ref{tableresults}. For the first (shorter) measurement, the longitudinal field has a relatively large error bar, but the next 3 (longer) measurements all show longitudinal fields of the order of -6 to -10 G. Similar measurements ($N_l$) were also performed with the N profiles and are compatible with 0.

\section{Discussion}\label{discus}

\subsection{Magnetic detections}

We confirm that the B star $\epsilon$\,CMa is magnetic, as discovered by
\cite{fossati2015}. However, we find that it is located at the end of the MS and
it is not yet a post-MS object, as already suspected by \cite{fossati2015}. It is nevertheless classified as a bright giant (B1.5II) in
the literature, likely due to its \ion{He}{ii} line. This is frequently the case for massive
stars in the second part of the MS, as shown by \cite{martins2017}. In the same
way, the O star $\zeta$\,Ori\,Aa was confirmed to be magnetic by
\cite{blazere2015}, but its small radius \citep[$\sim$20
R$_\odot$,][]{hummel2013} and its position on evolutionary tracks \citep{fossati2015} suggest that it is a very young supergiant 
still on the MS. Until now, there were thus no known clearly evolved (post-MS) magnetic
hot supergiants. 

In this work, we find that $\iota$\,Car and HR\,3890 are two magnetic A7Ib
supergiants. Evolutionary models presented in this work leave no doubt about
their strongly evolved supergiant status, well beyond the MS, and the magnetic signatures in the
stars are clear.

We recall that the evolutionary models used in this work do not include a magnetic field. It is known that a magnetic field can modify the evolution of the star. In particular, it can brake the stellar rotation, reduce its effective mass loss, and also modify its internal structure and rotation profile. Without a magnetic stellar evolution code, it is not possible to provide quantitative estimates of how the magnetic field did and will modify the evolution of the stars we have studied here. Our results, however, provide observational constraints to future evolutionary models with fossil magnetic fields that are currently under development.

Since our determination of the current evolutionary status of the targets studied here is based on their spectra and not on the models, the current evolutionary status presented above is certain. However, the stellar parameters derived for the stars at the ZAMS could be over- or under-estimated because of the lack of magnetic fields in the evolutionary models.

\subsection{Origin of the magnetic field in hot supergiants}

Following the two detections of magnetic fields in supergiant stars presented in
this work, the possible  origin of these fields must be discussed. With this aim, 
we computed Kippenhahn diagrams for both stars (see Fig.~\ref{Fig_Kippenhahn}) from the evolutionary models presented above.
They provide the evolution (without a magnetic field) of the stellar internal structure as a function of
time relative to the last computed time step ($t_{\rm f}$). The respective
radial extension of convective and stably stratified radiative regions are
represented in gray and white, respectively, while the red line indicates the stellar surface. In the case of $\iota$\,Car, the
star is currently in a transition phase: a convective external region is
developing above its large radiative envelope and a small convective core. In
the case of HR\,3890, the star has already passed this transition phase but only a very thin convective layer has developed in the radiative envelope.

Since the external layers of both stars are currently still mostly radiative, and have
remained mainly radiative during the entire post-MS evolution so far, the most
plausible hypothesis is that the current magnetic field is the same field as
those observed in B-type stars on the MS, i.e. a fossil field
\citep{BraithwaiteNordlund2006,DuezMathis2010}, which has been conserved. To
verify this hypothesis, we can use magnetic flux conservation to infer the
strength of the fossil field the stars would have hosted during their MS stage, using the stars' current and MS radius estimates. We obtain a MS
field between $\sim$700 G and $\sim$1100 G for $\iota$\,Car and between $\sim$3 and $\sim$6 kG
for HR\,3890, respectively (see Table~\ref{tableresults}, which also indicates a MS field for $\epsilon$\,CMa of about 200 G). These values correspond to the typical fossil field
strength observed in magnetic Bp or Ap stars on the MS 
\citep[e.g.][]{grunhutneiner2015}, therefore magnetic flux conservation appears
to be plausible. This does not imply that magnetic flux conservation alone has 
occurred, but simply that it is a sufficient explanation. If these magnetic
supergiants started their lives with a stronger magnetic field,
magnetic decay could also have occurred, as observed on the MS \citep{bagnulo2006,landstreet2007,landstreet2008,fossati2016}. In addition, the radius on the MS may be slightly larger/smaller when including a magnetic field in the evolutionary models \citep{duez2010}, hence the MS field might be slightly different as well.

In spite of the agreement between the derived MS field for our targets and typical fossil field strengths observed in magnetic Bp or Ap stars on the MS, we investigated the possible role of the central convective core
of each star. First, they have a very small radius (of the order of
0.1 R$_{\odot}$) when compared to the total radius of the star (of the order
of 100 R$_{\odot}$). Therefore, even if a dynamo action is at work in the core 
\citep{BrunBrowningToomre2005}, the field generated in these layers would have a
large distance to cross to reach the surface of the star. In this framework,
\cite{MacGregorCassinelli2003} demonstrated that the time needed for a flux tube
generated in the core of a massive MS star to emerge at the stellar
surface is longer than the lifetime of the star. This time is even longer in a
supergiant star. Therefore, we conclude that it is not possible that the fields
observed at the surface are those generated in the convective core. 

A dynamo could also develop in the intermediate thin convective layer of HR\,3890. The same argument as for the convective core however applies: the layer is very thin and located too deep (in the middle of the radiative envelope) for a potential dynamo field to reach the surface. According to our models, $\iota$\,Car did not reach the phase where a large external convective zone develops yet, and its Zeeman signature looks simple, as expected from a dipolar field, rather than complex, as expected from a dynamo field in a convective shell. 

Finally, the impact of the intermediate thin convective layer of HR\,3890 on its
fossil field is examined. Because of the small radial extent of this convective
layer relatively to the global size of the star, we may infer that it has not significantly perturbed the fossil field configuration in the external radiative
envelope during the star's evolution. Indeed, a similar situation occurs earlier
in the life of the star, when it transitions from the pre-MS Herbig phase to the
MS. During this transition, the appearance of a convective core does not affect the
fossil field in the radiative region \citep{Alecianetal2013}. However, it is 
possible that the obliquity of the fossil field changes, because of its
coupling with the dynamo fields likely present in the core and in the
intermediate convective shell \citep{Featherstoneetal2009}. 

From the above discussion, we conclude that the magnetic fields observed at the
surface of $\iota$\,Car and HR\,3890 are probably the remnants of the fields
present on the MS, i.e. they are fossil fields.

\section{Conclusions}\label{conclu}

In this paper we confirm that $\epsilon$\,CMa is magnetic, as discovered by \cite{fossati2015}, but we find that it is near the end of the MS rather than a post-MS star. In addition, we present the detections of two new magnetic A7 supergiant stars, which are more evolved: $\iota$\,Car and HR\,3890. Their current weak field is compatible with the strength of fossil fields observed on the MS if we consider magnetic flux conservation as the principal agent driving evolution of their surface magnetic field strengths. Moreover, the shape and slow evolution with time of their Zeeman signatures also point towards fossil fields. Therefore, these magnetic A supergiants are the probable descendants of magnetic MS B stars. Monitoring of their field variability over a full rotation period with further spectropolarimetric observations will allow us to verify the fossil origin of these fields and model their geometry and strength.

In the future it will also be interesting to find magnetic hot supergiants that are even more evolved and with a large external convective region that has developed on top of the radiative envelope (i.e. a hotter, more evolved  version of $\iota$\,Car). Their fossil fields should be even weaker than those presented here, and those fields could possibly be more complex if they are mixed with dynamo effects developing in the convective surface envelope \citep{augustson2017}. Their observation will thus require ultra-deep spectropolarimetric observations.

Finally, the results presented here provide observational constraints for the current development of evolutionary models of hot stars including a fossil magnetic field.

\section*{Acknowledgements}

CN and AB acknowledge support from the ANR (Agence Nationale de la Recherche)
project Imagine. CN, MO, SM, AB, and BB acknowledge support from PNPS (Programme National de Physique Stellaire). MK acknowledges support by the Austrian Science Fund FWF within the DK-ALM (W1259-N27).
Part of the research leading to these results has received funding from the European Research Council (ERC) under the European Union's Horizon 2020
research and innovation programme (grant agreements number 670519: MAMSIE and 647383: SPIRE). GAW acknowledges Discovery Grant support from the Natural Sciences and Engineering Research Council (NSERC) of Canada. This research has made use of the SIMBAD database operated at
CDS, Strasbourg (France), and of NASA's Astrophysics Data System (ADS). 


\bibliographystyle{mnras}
\bibliography{articles} 


\bsp	
\label{lastpage}
\end{document}